\documentclass[fullpages,toclabels,10pt]{article}

\usepackage{epsfig,psfrag,subfig}
\usepackage{amsmath,amsthm,amssymb,amsfonts}
\usepackage{arydshln}
\usepackage{enumerate}
\usepackage{KKKKSymb}
\usepackage{framed}

\newtheorem{problem}{\bf Problem}

\newcommand{\opt}{{\star}}
\newcommand{\vb}{{\bf v}}
\newcommand{\C}{\mathbb{C}}

\allowdisplaybreaks

\usepackage{hyperref}
\hypersetup{
    colorlinks=true,%
    citecolor=blue,%
    filecolor=black,%
    linkcolor=red,%
    urlcolor=black
}

\usepackage[on]{auto-pst-pdf}

\usepackage{color}

\title{
\sc Mathematical Programs for Belief Propagation and Consensus
}

\author{\\
        {\bf Kwang-Ki K. Kim} \\
        School of Electrical and Computer Engineering\\
        Georgia Institute of Technology\\
        Atlanta, Georgia {USA} \\
        {\tt kwangki.kim@ece.gatech.edu}\\[5mm]
}
\date{\today}

\begin{document}
\maketitle
\tableofcontents

\begin{abstract}

This paper develops methods of distributed Bayesian hypothesis tests for fault detection and diagnosis that are based on belief propagation and optimization in graphical models.
The main challenges in developing distributed statistical estimation algorithms are
i) difficulties in ensuring convergence and consensus for solutions of distributed inference problems,
ii) increasing computational costs due to lack of scalability, 
and
iii) communication constraints for networked multi-agent systems.
To cope with those challenges, this manuscript considers
i) belief propagation and optimization in graphical models of complex distributed systems,
ii) decomposition methods of optimization for parallel and iterative computations,
and
iii) distributed decision-making protocols.

\end{abstract}

\section{Introduction}
\label{sec:introduction}

Stochastic inference using graphical models \cite{cowell1998introduction,wainwright2008graphical} have been significantly important research topics in a variety of disciplines that include signal processing \cite{sudderth2004embedded}, machine learning \cite{frey1998graphical}, and artificial intelligence \cite{pearl1988probabilistic}.
For the use of graphical models in statistical inference problems, optimal fusion of information and/or data over networked agents that are individual decision makers or processors and the design of compromised inference methods for distributed decision makers have far significant importance. 

A monumental work of Pearl~\cite{pearl1988probabilistic} called {\it belief propagation} (BP) is a message-passing algorithm for which local evidences are exchanged as messages that are used to update local beliefs and to find fixed-points of iterations, corresponding to marginal probability distributions of the node states.
In a standard BP method for statistical inference in a graphical model, agents on the nodes exchange messages with neighboring agents connected over the edges.
The BP algorithm is known to provide exact marginal distributions when the graphical model are tree-structured, i.e., of no cyclic loops~\cite{pearl1988probabilistic}.
In the presence of cyclic loops in a graphical model, neither convergence nor optimally of BP methods cannot be, in general, guaranteed, whereas some empirical studies on performance of loopy BP~\cite{murphy1999loopy} and conversion to equivalent cycle-free graphical models~\cite{cowell1998advanced} are available.  

The main challenges in the development of BP algorithms with general Markov and Bayesian graphical models are
\begin{enumerate}[i)]
\item
{\bf Convergence Analysis}:
As we previously mentioned, message-passing algorithms of BP do not generally converge to a fixed-point in the presence of cyclic loops.
\item
{\bf Scalability}:
In a tree-structured graphical model, BP algorithms can find a fixed-point in $O(n)$ iterations where $n$ is the diameter of the graph.
However, calculation of posterior marginal probabilities on nodes in an arbitrary bayesian network is known to be NP-complete~\cite{cooper1990computational,shimony1994finding}
and even an approximate computation of posterior marginal probabilities is NP-hard~\cite{dagum1993approximating}.
\item
{\bf Communication Constraints}:
Message-passing or information-exchange over communication networks are not necessarily reliable, and communication bandwidth and energy constraints are typical sources of degrading performance of networked inference algorithms~\cite{cetin2006distributed}.
\end{enumerate}

To cope with the aforementioned difficulties confronted to BP methods for statistical inference in graphical models, we consider
\begin{enumerate}[$\rightarrow$i)]
\item
{\bf Belief Optimization}:
In \cite{yedidia2005constructing}, it was shown that BP fixed-points correspond to the stationary points of the Bethe free energy approximation for a factor graph.
The associated constrained minimization is called {\it belief optimization} (BO).
Our statistical inference methods are based on the same principle that the joint probability distribution of the node states in a graphical model is a minimizer of the free energy and the beliefs, corresponding to marginal probabilities of the node states, can be computed from minimizing approximate free energy such the mean field and Bethe free energies.
The resultant statistical inference problems are given as constrained minimization.
\item
{\bf Decomposition Methods of Optimization}:
Belief optimization is large-scale constrained minimization that becomes intractable and non-scalable as the number of nodes and cardinality of the node states increase.
Since the coupling between marginal probabilities to be determined are constrained on the edges in graphical models,
natural ways of reducing computational demand are to use decomposition methods for optimization.
\item
{\bf Distributed Decision Processes}:
In the presence of communication constraints, decision processes and information exchange need to be localized and distributed for reliable statistical inference over graphical models.
\end{enumerate}

Our main applications of BP/BO methods are distributed hypothesis tests for fault detection and diagnosis (FDD) in large-scale distributed dynamical systems. 
Developing automatic monitoring, detection, and diagnosis of system faults has rapidly growing importance as the size and complexity of systems increase.
Most of existing methods for model-based FDD are centralized schemes in the sense that the central decision maker can access all measurements
and the decision goal is to decide whether faults occur and determine types and locations of faults.  
Distributed FDD is suitable for large-scale interconnected and networked dynamical systems such as multi-agent systems and power grids.
Furthermore, since not all measurements are accessible to local processors and computation nodes, centralized FDD schemes may not be applicable to distributed systems.
Belief propagation and optimization provide naturally suitable ways of distributed statistical inference and decision making,
for which graphical models are used for representation of interconnections and networks of local sensors (measurements) and processors (data/information-processing)
and belief consensus constraints are required to be satisfied by exchanging messages for BP and by imposing public variable constraints for BO.


\section{Belief Propagation in Graphical Models}
\label{sec:}

BP algorithms are developed for graphical models.
This section provides a concise discussion of graphical representations and the corresponding BP methods for distributed inference problems. 
There are two types of graphical models that are used to represent probabilistic and informational dependencies of random variables--Markov networks and Bayesian networks.
A Markov network is defined with an undirected graph whose nodes correspond to random variables and the edges correspond to their probabilistic and information dependencies.
A Bayesian network is defined with a directed graph whose nodes correspond to random variables and the arrows are used to denote causality constraints or class-property relations.
Since our focus is on developing distributed Bayesian hypothesis tests for FDD using BP/BO, we only consider Markov network models. 
Many research monographs for tutorial of graphical models are available (see \cite{cowell1998introduction,clifford1990markov,wainwright2008graphical}, for example).

\subsection{Pairwise MRF}

Markov networks (aka Markov random field (MRF) models) are suited for representing conditional dependencies of the node states. 
\begin{definition}[MRF]
The random vector $X$ is Markov with respect to the graph $G = (V,E)$ if for any partition of the node set $V$ into disjoint sets $A$, $B$, $C$, in which $B$ separates $A$ and $C$,
the degenerate random vectors $X_A$, $X_B$, $X_C$ corresponding to each node set are conditionally independent in the sense that $P_{AB|C}(x_a, x_b | x_c) = P_{A|B}(x_a| x_b) P_{C|B}(x_c| x_b)$, or equivalently $P_{A|BC}(x_a | x_b, x_c) = P_{A|B} (x_a|x_b)$ (or symmetrically, $P_{C|AB}(x_c | x_a, x_b) = P_{C|B} (x_c|x_b)$).
\end{definition}

The next theorem called The HammersleyÐClifford theorem provides a sufficient (and necessary) condition for which the joint probability distribution of the node states can be represented as an MRF. 
\begin{theorem}[Hammersley-Clifford Theorem (see \cite{clifford1990markov,pearl1988probabilistic})]
The random vector $X$ is Markov w.r.t. the graph $G$ if (and only if for strictly positive probability distributions) its distribution can be factorized by a product of variables restricted to cliques, i.e., the joint probability can be factorized as the followings:
\begin{equation}	\label{eq:clique_fact}	
P(\x) = \gamma \prod_{C \in \Ccal} \psi_{C}(x_c)
\end{equation}
where $\gamma = (\sum_{x} \prod_{C \in \Ccal} \psi_{C} (x_c))^{-1}$ and $\Ccal$ refers to the set of cliques in $G$.
\end{theorem}
The $\psi_{C}(x_c)$ are called the {\it compatibility functions} that correspond to the marginal probabilities, and their negative logarithms are referred to as potentials or potential functions,
$V_C(x_c) : = -\ln \psi_{C}(x_c) \geq 0$.
The factorization \eqref{eq:clique_fact} can be rewritten as
\begin{equation}	\label{eq:clique_fact:2}	
P(\x) = \gamma \left( \prod_{k \in V} \psi_{k}(x_k) \right) \left( \prod_{(i,j) \in E} \psi_{ij} (x_i,x_j) \right) \left( \prod_{C \in \Ccal \setminus V, E}  \psi_{C}(x_c) \right).
\end{equation}
\begin{assumption}[Pairwise Potentials]	\label{ass:pairwiseMRF}
We assume that either 
\begin{enumerate}[i.]
\item
there is no clique with more than two nodes in the graph $G$, or
\item
the potentials are only defined by the variable as a single node in $V$ or by the two variables as a pair of nodes on an edge in $E$.
\end{enumerate}
\end{assumption}
Under Assumption \ref{ass:pairwiseMRF}, there is no contribution of the last term in \eqref{eq:clique_fact:2}, i.e., 
\begin{equation}	\label{eq:clique_fact:approx}	
P(\x) \equiv \hat{P}(\x) \triangleq  \gamma \left( \prod_{k \in V} \psi_{k}(x_k) \right) \left( \prod_{(i,j) \in E} \psi_{ij} (x_i,x_j) \right)
\end{equation}
where $\hat{P}(\x)$ can be interpreted as an approximation of the joint probability distribution $P(x)$ of the random variable $X$ that is Markov w.r.t. $G = (E,V)$, up to the $2$-cliques.

\subsubsection{Graphical models for distributed statistical inference}

From here, we assume that there are local measurements (or evidences) $y_k \in \Ycal_k$ that are associated with the node $k \in V$.
For any non-loopy graph, i.e, graphical models on trees, the compatibility functions can be represented in terms of the marginal probabilities up to the $2$-cliques:
$\psi_{k}(x_k) = p_{k}(x_k) p(y_k | x_k)$ for $k \in V$ and  $\psi_{ij} (x_i,x_j) = p_{ij}(x_i,x_j)p(y_i,y_j | x_i,x_j) /p_{i}(x_i)p(y_i | x_i) p_{j}(x_j)p(y_j | x_j)$ for $(i,j) \in E$.
With this representation of the compatibility functions, $\hat{P}(X)$ can be rewritten as
\begin{equation}	\label{eq:clique_fact:approx:2}	
\hat{P}(\x)=  \gamma \left( \prod_{k \in V} p_{k}(x_k) p(y_k | x_k) \right) \left( \prod_{(i,j) \in E} \frac{p_{ij}(x_i,x_j)p(y_i,y_j | x_i,x_j) }{p_{i}(x_i)p(y_i | x_i)  p_{j}(x_j)p(y_j | x_j)} \right)
\end{equation}
or
\begin{equation}	\label{eq:clique_fact:approx:3}	
\hat{P}(\x)=  \gamma \left( \prod_{k \in V}  p(x_k | y_k) \right) \left( \prod_{(i,j) \in E} \frac{p(x_i,x_j | y_i,y_j) }{p(x_i | y_i) p(x_j | y_j)} \right)
\end{equation}
where, for abusing notation, $\gamma$ might not be the same as the one in \eqref{eq:clique_fact:approx}, but can be considered as an equivalent partition function (value).

For the purpose of distributed statistical inference in a graphical model, a goal is to estimate the posterior marginal probabilities,
for which messages from the neighboring nodes are required to have sufficient statistics of local measurements that can be considered as realizations from unknown probability distributions.
\begin{problem}	\label{prob:marginal-probabilities}
Consider an undirected graph $G= (V,E)$.
Compute (or approximate) the posterior marginal probabilities
\begin{equation}
p_k(x_k| y_1, \cdots, y_N), \quad k \in V
\end{equation}
where $N = |V|$.
\end{problem}
To exactly solve Problem \ref{prob:marginal-probabilities}, the required property of a BP method is the relation of sufficient statistics
\begin{equation}
p_k ( x_k | y_1, m_k )   \equiv p_k(x_k| y_1, \cdots, y_N), \quad k \in V
\end{equation}
where $m_k$ refers to the total messages delivered to Agent at the node $k$.\footnote{Agent $k$ refers to a processor or decision maker at the node $k$.}

\subsubsection{Distributed belief propagation}

\begin{algorithm}[Belief propagation algorithm]
In a belief propagation algorithm, 
the belief at the node $k$ in its state $x_k$ is 
\begin{equation}
\beta_{k} (x_k) \propto \psi_k (x_k) \prod_{\ell \in \Ncal(k)} \mu_{\ell \to k} (x_k)
\end{equation}
and the message from the node $\ell$ to the node $k$ about the state $x_k$ can be either the sum-product BP message
\begin{equation}
\mu_{\ell \to k} (x_k) \propto 
\sum_{x_{\ell}} \psi_{\ell k}(x_{\ell}, x_{k}) \psi_{\ell}(x_{\ell}) \prod_{u \in \Ncal(\ell) \setminus \{k\}} \mu_{u \to \ell} (x_{\ell})
\end{equation}
or the max-product BP message
\begin{equation}
\mu_{\ell \to k} (x_k) \propto 
\max_{x_{\ell}} \psi_{\ell k}(x_{\ell}, x_{k}) \psi_{\ell}(x_{\ell}) \prod_{u \in \Ncal(\ell) \setminus \{k\}} \mu_{u \to \ell} (x_{\ell}),
\end{equation}
where conditional dependence of the beliefs and messages on measurements $Y= \{y_i\}_{i=1}^{n}$ is dropped for the sake of notation.
\end{algorithm}

In the aforementioned BP algorithms, there are slightly different methods of computing messages to be transmitted.
They have different interests \cite{yedidia2005constructing}:
(a) the max-product BP message is to obtain a global state that is most probable in the Bayesian sense
and consists of a local state maximizing the local belief, and
(b) the sum-product BP message is to compute marginal posterior probabilities, given the total evidence or measurements that are available in the system.
Their properties need to be clarified.

\paragraph{The Max-Product BP}

A goal of a belief propagation algorithm for Bayesian estimation, particularly for a maximum a posteriori estimation, can be to achieve the relation
\begin{equation}
\beta_{k}(x_k) = \alpha_k \max_{x_{-k}} p_k \left( x_k, x_{-k} | y_1, \cdots, y_N \right), \quad \forall x_{k}, \ \forall k \in V,
\end{equation}
for given total measurement data $\{y_{k}\} \in Y$, where each $\alpha_k$ is a positive constant that is independent of the value of $x_k$ and results in $\beta_k (\cdot) \in [0,1]$.
Alternatively, a slightly weaker relation is that for given measurement data $\{y_{k}\}$,
\begin{equation}
\beta_{k}(x) \leq \beta_{k} (z) \ \Rightarrow \  \max_{x_{-k}} p_k \left( x , x_{-k} | y_1, \cdots, y_N \right) \leq  \max_{x_{-k}} p_k \left( z , x_{-k} | y_1, \cdots, y_N \right),
\end{equation}
for all nodes $k \in V$.
Note that the above relation can ensure the marginal maximum a posteriori (m-MAP) estimation, i.e., 
\begin{equation}
\begin{split}
x_k^{\opt} 
& =
\arg \max_{x} \beta_{k} (x) \\
& = 
\arg \max_{x} p_k \left( x, x_{-k}^{\opt} | y_1, \cdots, y_N \right)
\end{split}
\end{equation}
and they indeed result in the joint MAP (j-MAP) estimator satisfying the relation
\begin{equation}
\{x_k^{\opt}\} = \arg \max_{\{x_i\}} p \left( x_1, \cdots, x_N | y_1, \cdots, y_N \right).
\end{equation}
%

\paragraph{The Sum-Product BP}

Similar to the max-product BP algorithm, the goal of the sum-product BP is to achieve the relation
\begin{equation}
\beta_{k}(x_k) = \alpha_k  \sum_{x_{-k}} p_k(x_k,x_{-k} | y_1, \cdots, y_N ), \ \forall k \in V,
\end{equation}
where the summation is computed for all realizations of the compound random vector $x_{-k}$ and each $\alpha_k$ is a positive constant that is independent of the value of $x_k$ and results in $\beta_k (\cdot) \in [0,1]$.
Note that this is indeed to estimate the marginal posterior probabilities, for given total measurements.

\begin{remark}
A notable discrimination of the sum-product BP against the max-product BP is that the combination of optimal m-MAP estimators
$x_k^{\opt} = \arg \max_{x} \beta_{k} (x)$, where the beliefs are obtained from the sum-product BP,
does not necessarily compose of an optimal j-MAP estimation.
\end{remark}

\paragraph{Iterative Message-Passing and Fixed-Points}

The following algorithm is a standard asynchronous iterative message-passing algorithm for belief propagation. 
\begin{algorithm}[Parallel iterative message-passing algorithm]
The belief at the node $k$ in its state $x_k$ at time $t$ is 
\begin{equation}
\beta_{k}^{(t)} (x_k) \propto \psi_k (x_k) \prod_{\ell \in \Ncal(k)} \mu_{\ell \to k}^{(t)} (x_k)
\end{equation}
and the message from the node $\ell$ to the node $k$ about the state $x_k$ at time $t$ can be either the sum-product BP message update
\begin{equation}
\mu_{\ell \to k}^{(t)} (x_k) \propto 
\sum_{x_{\ell}} \psi_{\ell k}(x_{\ell}, x_{k}) \psi_{\ell}(x_{\ell}) \prod_{u \in \Ncal(\ell) \setminus \{k\}} \mu_{u \to \ell}^{(t-1)} (x_{\ell})
\end{equation}
or
the max-product BP message update
\begin{equation}
\mu_{\ell \to k}^{(t)} (x_k) \propto 
\max_{x_{\ell}} \psi_{\ell k}(x_{\ell}, x_{k}) \psi_{\ell}(x_{\ell}) \prod_{u \in \Ncal(\ell) \setminus \{k\}} \mu_{u \to \ell}^{(t-1)} (x_{\ell}).
\end{equation}
\end{algorithm}



\section{Belief Optimization in Graphical Models}
\label{sec:}

\subsection{Bethe-Peirerls Approximation to the Free Energy}

In~\cite{yedidia2001bethe,yedidia2003understanding,yedidia2005constructing}, the authors showed that the fixed points of BP and its generalization are indeed associated with extrema of the Bethe and Kikuchi free energies, respectively. 
Here, we provide a concise overview of some useful results from statistical physics.
In particular, the observation that statistical inference problems can be represented as minimization of (approximate) free energy (see also~\cite{yedidia2001bethe,yedidia2005constructing})
motivates to study various approximate free energy.

\subsubsection{Gibbs free energy in statistical physics}

In statistical physics, the Boltzmann distribution law tells us that for the energy $E(\x)$ associated with some state or condition $\x$ of a system, the probability distribution of its occurrence
is given by 
\begin{equation}
p(\x) = \frac{1}{Z} \exp (-E(\x)/T)
\end{equation}
where $Z$ denotes the partition function (constant) and $T$ is the temperature that can be set to be $1$ without loss of generality.
Comparing this to the factorization \eqref{eq:clique_fact} gives
$\gamma = 1/Z$ and $E(x) =- \sum_{C \in \Ccal} \ln \psi_{C}(x_c) =  \sum_{C \in \Ccal}  V_{C}(x_c)$, i.e., the total energy is the sum of potentials over the system. 
To compute the distance between the belief $\beta(\x)$ and the true joint probability distribution, use the Kullback-Leibler (KL) distance that is defined by 
\begin{equation}
\begin{split}
D(\beta || p) 
& = \sum_{\x} \beta(\x) \ln \frac{\beta(\x)}{p(\x)}  \\
& = \sum_{\x}  \beta(\x) E(\x) + \sum_{\x}\beta(\x) \ln \beta(\x) + \ln Z
\end{split}
\end{equation}
such that $D(\beta || p) =0$ if and only if $\beta \equiv p$ and $D(\beta || p) \geq 0$ for all $\beta \in \Delta$ where $\Delta$ refers to the set of probabilities.
Define the Gibbs free energy by 
\begin{equation}
G(\beta) \triangleq  \sum_{\x}  \beta(\x) E(\x) + \sum_{\x}\beta(\x) \ln \beta(\x) = U(\beta) - H(\beta)
\end{equation}
such that $D(\beta || p) = G(\beta) - F$ where $F \triangleq -\ln Z$ is called the Helmholtz free energy, and
$U(\beta)$ and $H(\beta)$ refer to the average energy and the entropy, respectively.

\subsubsection{Approximate free energy}

Previously, we assumed that the joint probability $p(\x)$ is a function of the total energy function $E(\x)$.
Suppose that the system is of a pairwise MRF with the graph $G(V,E)$ in which there is no potential related to cliques with more than two nodes.
Then the corresponding energy of such a configuration is
\begin{equation}
E(\x) = - \sum_{k \in V} \ln \psi_{k} (x_k) - \sum_{(i,j) \in E} \ln \psi_{ij}(x_i,x_j) .
\end{equation}
%

\paragraph{\it A. The Mean Field Free Energy}

In the mean-field theory, the joint distribution $\beta(\x)$ is approximated by the complete factorization, i.e,
\begin{equation}	\label{eq:mean-field:factorization}
\beta(\x) \approx \prod_{k \in V} \beta_k ( x_k).
\end{equation}
With this approximate joint distribution under a pairwise MRF configuration, the mean-field average energy is
\begin{equation}	\label{eq:mean-field:average-energy}
\tilde{U}(\{ \beta_{\ell} \}_{\ell \in V}) = - \sum_{k \in V} \sum_{x_k} \beta_{k} (x_k) \ln \psi_{k} (x_k) - \sum_{(i,j) \in E} \sum_{x_i,x_j}  \beta_{i} (x_i) \beta_{j} (x_j) \ln \psi_{ij}(x_i,x_j)
\end{equation}
and similarly the mean-field entropy is
\begin{equation}	\label{eq:mean-field:entropy}
\tilde{H} (\{ \beta_{\ell} \}_{\ell \in V} ) = - \sum_{k \in V} \sum_{x_k} \beta_{k} (x_k) \ln \beta_{k} (x_k).
\end{equation}
Note that the mean field free energy $\tilde{G} = \tilde{U} - \tilde{H}$ is a function of the separate one-node beliefs $\beta_k(\cdot)$.

\paragraph{\it B. The Bethe Free Energy}

For more general approximation, the joint distribution $\beta(\x)$ can be approximated by the factorization with one- and two-nodes beliefs, viz,
\begin{equation}	\label{eq:bethe:factorization}
\beta(\x) \approx \frac{\prod_{(i,j) \in E} \beta_{ij} ( x_i, x_j)}{\prod_{k \in V} \beta_{k}(x_k)^{q_k-1}}
\end{equation}
where $q_k = |\Ncal(k)|$.
With this approximate joint distribution under a pairwise MRF configuration, the Bethe average energy is
\begin{equation}	\label{eq:bethe:average-energy}
\begin{split}
\tilde{U}(\{ \beta_{k} \}_{k \in V}, \{ \beta_{ij} \}_{(i,j) \in E}) 
= & - \sum_{k \in V} \sum_{x_k} \beta_{k} (x_k) \ln \psi_{k} (x_k) \\
    & - \sum_{(i,j) \in E} \sum_{x_i,x_j}   \beta_{ij} (x_i, x_j) \ln \psi_{ij}(x_i, x_j)
\end{split}
\end{equation}
and similarly the Bethe entropy is
\begin{equation}	\label{eq:bethe:entropy}
\begin{split}
\tilde{H} (\{ \beta_{k} \}_{k \in V}, \{ \beta_{ij} \}_{(i,j) \in E}) 
= & \sum_{k \in V} (q_k - 1) \sum_{x_k} \beta_{k} (x_k) \ln \beta_{k} (x_k) \\
  & - \sum_{(i,j) \in E} \sum_{x_i,x_j} \beta_{ij} (x_i, x_j) \ln \beta_{ij} (x_i, x_j).
\end{split}
\end{equation}
\begin{remark}
In contrast to the mean-field energy, the Bethe free energy is not generally an upper bound on the true Gibbs free energy~\cite{yedidia2005constructing}.
\end{remark}


\subsection{Belief Optimization}
\label{subsec:belief-opt}

\newcommand{\betav}{\underline{\beta}}
\newcommand{\betavk}{\underline{\beta_k}}
\newcommand{\psivk}{\underline{\psi_k}}
\newcommand{\betavi}{\underline{\beta_i}}
\newcommand{\psivi}{\underline{\psi_i}}
\newcommand{\betavj}{\underline{\beta_j}}
\newcommand{\psivj}{\underline{\psi_j}}
\newcommand{\betavell}{\underline{\beta_{\ell}}}
\newcommand{\betamij}{\underline{\underline{\beta_{ij}}}}
\newcommand{\psimij}{\underline{\underline{\psi_{ij}}}}

Consider the discrete random variables $X_k \in \Xcal_k \triangleq \{ x_{k1}, x_{k2}, \cdots, x_{kn_k}\}$ with probability one and $|\Xcal_k| = n_k$ for each $k \in V$.
For the sake of notation, assume that all the nodes have the same cardinality of their supports, i.e. , $n_k = n$ for all $k \in V$.
Define the probability vector and matrix by
\begin{equation}
\betavk \triangleq
\left[ 
\begin{array}{@{}c@{}}
\beta_{k}(x_{ki}) \\
\vdots \\
\beta_{k}(x_{kn}) 
\end{array}
\right]
, \ 
\mbox{for } k \in V
\end{equation}
and
\begin{equation}
\betamij \triangleq
\left[ 
\begin{array}{@{}ccc@{}}
\beta_{ij}(x_{i1}, x_{j1}) & \cdots & \beta_{ij}(x_{i1}, x_{jn})  \\
\vdots & \ddots & \vdots \\
\beta_{ij}(x_{in}, x_{j1}) & \cdots & \beta_{ij}(x_{in}, x_{jn})  
\end{array}
\right]
, \ 
\mbox{for } (i,j) \in E,
\end{equation}
respectively.
The Belief Optimization (BP) is to find $\{\betavk\}_{k \in V}$ minimizing $\tilde{G}$ for the mean-filed free energy approximation or
$( \{\betavk\}_{k \in V}, \{\betamij\}_{(i,j) \in E} )$ minimizing $\tilde{G}$ for the Bethe free energy approximation.

\subsubsection{Minimization of the mean field free energy}

A popular method of approximating a free energy is the aforementioned mean field approach for which an optimal configuration of beliefs, that is an approximation of joint probability distribution,
can be obtained as a factorization \eqref{eq:mean-field:factorization} and the associated factors $\{\betavk\}_{k \in V}$ are optimal solutions of the constrained minimization
\begin{equation}
\begin{split}
\min \ \ &  \tilde{G} = \tilde{U} - \tilde{H} \\
\mbox{s.t.} \ \  & e^{\T} \betavk = 1, \ k \in V \\
				& 0 \leq \betavk \leq 1, \ k \in V
\end{split}
\end{equation}
where $\tilde{U}$ and $\tilde{H}$ are given by \eqref{eq:mean-field:average-energy} and \eqref{eq:mean-field:entropy}, respectively.
It can be explicitly rewritten as
\begin{equation}	\label{opt:mean-field-energy-min}
\begin{split}
\min \ \ &  - \sum_{k \in V} \betavk^{\T} \ln \psivk - \sum_{(i,j) \in E} \betavi^{\T} \ln \psimij \, \betavj + \sum_{k \in V} \betavk^{\T} \ln \betavk \\
\mbox{s.t.} \ \  & \betavk \in \Delta,\,  k \in V \\
\end{split}
\end{equation}
where $\Delta \triangleq \{ p \in \Real^{n} :  e^{\T} p = 1, \, p_i \in [0,1], \, \forall i \}$.

\subsubsection{Minimization of the Bethe free energy}

Similar to minimization of the mean field free energy, an optimal configuration of beliefs, that is an approximation of joint probability distribution,
can be obtained as a factorization \eqref{eq:bethe:factorization} and the associated factors $( \{\betavk\}_{k \in V}, \{\betamij\}_{(i,j) \in E} )$ are optimal solutions of the constrained minimization
\begin{equation}
\begin{split}
\min \ \ &  \tilde{G} = \tilde{U} - \tilde{H} \\
\mbox{s.t.}\ \ & e^{\T} \betavk = 1, \ k \in V \\
				& 0 \leq \betavk \leq 1, \ k \in V \\
				& e^{\T} \betamij = \betavj^{\T}, \, \betamij e = \betavi,\, (i,j) \in E
\end{split}
\end{equation}
where $\tilde{U}$ and $\tilde{H}$ are given by \eqref{eq:bethe:average-energy} and \eqref{eq:bethe:entropy}, respectively.
It can be explicitly rewritten as
\begin{equation}	\label{opt:bethe-energy-min}
\begin{split}
\min \ \
& - \sum_{k \in V} \betavk^{\T} \ln \psivk  - \sum_{(i,j) \in E} [\betamij \circ \ln \psimij] \\
& \ 
+ \sum_{k \in V} (1- q_k) \betavk^{\T} \ln \betavk + \sum_{(i,j) \in E} [\betamij \circ \ln \betamij]
\\
\mbox{s.t.} \ \  & \betavk \in \Delta,\,  k \in V \\
				& e^{\T} \betamij = \betavj^{\T}, \, \betamij e = \betavi,\, (i,j) \in E
\end{split}
\end{equation}
where $[A \circ B] = {\rm Tr} (A^{\T} B)$ refers to entry-wise sum of the Hadamard (aka Schur) product  $A \circ B$.

\subsubsection{Minimization of the TAP free energy}

The TAP (Thouless-Anderson-Palmer) approach that is used to approximate free energy in statistical mechanics has been adopted robust decoding and statistical inference
based on belief propagation (see \cite{saad1999belief,csato2001tap,kabashima2007belief}, for example).
an optimal configuration of beliefs, that is an approximation of joint probability distribution,
can be obtained as a factorization \eqref{eq:mean-field:factorization} and the associated factors $\{\betavk\}_{k \in V}$ are optimal solutions of the constrained minimization
\begin{equation}
\begin{split}
\min \ \ &  \tilde{G} = \tilde{U} - \tilde{H} - \tilde{T} \\
\mbox{s.t.} \ \  & \betavk \in \Delta,\,  k \in V 
\end{split}
\end{equation}
where $\tilde{T}$ refers to the TAP-correction to the mean field free energy.
%
%
This belief optimization based on the TAP free energy approximation is similar to the mean field free energy approach for which the marginal probability distributions are assumed to be independent.
In addition, the TAP free energy approach can be considered as an approximation of the Bethe free energy approach up to the second order moment~\cite{welling2001belief}.
Due to its similarity to the mean field energy approach and lack of accuracy, compared to the Bethe free energy approach,
we only focus on using the mean field and the Bethe free energy approaches and solving the corresponding constrained minimization problems.



\section{BP/BO Approaches to Belief Consensus}
\label{sec:belief-consensus:fdd-applications}

This section develops decomposed methods to solve the optimizations presented in Section \ref{subsec:belief-opt}.
In particular, methods of dual decomposition (see Appendix \ref{subsec:dual-decomp}) that solve the associated large-scale optimization are used for decentralized/distributed computations.

\subsection{Belief Consensus: Dual Decomposition Approaches}

\subsubsection{Minimization of the mean field free energy}

Consider the constrained minimization \eqref{opt:mean-field-energy-min}.
This large-scale optimization over a graphical model can be decomposed into separated constrained minimizations for which Agent $i$ solves the optimization
\begin{equation}	\label{opt:mean-field-energy-min:separated}
\begin{split}
\min_{\beta_{i},\, \{\beta_{j}\}_{j \in \Ncal(i)}} \ \ &  -  \betavi^{\T} \ln \psivi - \sum_{j \in \Ncal(i)} \betavi^{\T} \ln \psimij \, \betavj + \betavi^{\T} \ln \betavi \\
\mbox{s.t.} \ \  & \betavi \in \Delta, \\
				& \betavj =\betavi, \, \forall j \in \Ncal(i),  
\end{split}
\end{equation}
where the second constraint corresponds to the consensus between the agents on edges connecting the node of Agent $i$
and $\Ncal(i)$ refers to the set of Agents neighboring Agent $i$.

For fault detection and diagnosis with multiple hypotheses, assume that each agent has the same bank of hypothesized models and the objective of resultant distributed decision-making is to obtain optimal marginal beliefs $\{\betavi\}_{i \in V}$ that achieve the consistency in localized estimations, i.e.,
\begin{equation}	\label{eq:MFE:belief-consensus}
\mbox{\bf Marginal Belief Consensus I: }
\beta_{i}(x) = \beta (x), \ \forall x \in \Xcal_{i}, \, \forall i \in V
\end{equation}
which can be rewritten as 
\begin{equation}	\label{eq:MFE:belief-consensus:vec}
\betavi = \betav , \, \forall i \in V, \, \mbox{ for some } \betav \in \Delta.
\end{equation}
Incorporating the consensus requirement \eqref{eq:MFE:belief-consensus:vec} into \eqref{opt:mean-field-energy-min:separated} results in a decomposed optimization
for which Agent $i$ solves
\begin{equation}	\label{opt:mean-field-energy-min:separated:consensus}
\begin{split}
\min_{\betavi,\, \betav} \ \ &  -  \betavi^{\T} \ln \psivi + \betavi^{\T} \underline{\underline{M_{i}}} \betavi + \betavi^{\T} \ln \betavi \\
\mbox{s.t.} \ \  & \betavi = \betav \in \Delta,
\end{split}
\end{equation}
where $\underline{\underline{M_{i}}}  \triangleq - \sum_{j \in \Ncal(i)}  \ln \psimij$ are nonnegative matrices since their entries correspond to compatibility functions or constraints and can be normalized to be in the interval $[0,1]$ without deforming configuration of the free energy with respect to the beliefs.
Notice that the pseudo variable $\betav$ is a global variable that is required to be the same in all decomposed (slave) problems. 

\paragraph{Case 1: [For $\underline{\underline{M_{i}}}  \succeq 0$]}
If the pairwise compatibility matrix $\underline{\underline{M_{i}}}$ is positive semidefinite then the optimization \eqref{opt:mean-field-energy-min:separated:consensus} is convex
and can be solved by using iterative dual decomposition methods, for which computations are decentralized for each Agent $i$ and belief consensus is achieved by iterations to find an optimal Lagrange multipliers.
For details of the use of dual decomposition methods and underlying theories, see Appendix \ref{subsec:dual-decomp}.

\paragraph{Case 2: [For $\underline{\underline{M_{i}}}  \succeq_{\Delta} 0$]}
If the pairwise compatibility matrix $\underline{\underline{M_{i}}}$ is conditionally positive semidefinite over the standard simplex $\Delta$ then
the optimization \eqref{opt:mean-field-energy-min:separated:consensus} is convex.
However, checking if $\underline{\underline{M_{i}}}  \succeq_{\Delta} 0$ is indeed NP-hard \cite{Murty_Kabadi_1987:mp}.
If a prior knowledge of $\underline{\underline{M_{i}}}  \succeq_{\Delta} 0$ is available, then one can use the same dual decomposition methods as Case 1.
If there is no condition $\underline{\underline{M_{i}}}  \succeq_{\Delta} 0$ a priori, then one might use semidefinite programming relaxation that can be found in the subsequent Case 3. 

\paragraph{Case 3: [Indefinite $\underline{\underline{M_{i}}}$]}
The optimization \eqref{opt:mean-field-energy-min:separated:consensus} can be rewritten as
\begin{equation}	\label{opt:mean-field-energy-min:separated:consensus:re}
\begin{split}
\min_{\betavi,\, \betav,\, B_i} \ \ &  -  \betavi^{\T} \ln \psivi + \langle \underline{\underline{M_{i}}} , B_i \rangle + \betavi^{\T} \ln \betavi \\
\mbox{s.t.} \ \  & \betavi = \betav \in \Delta, \\
			&  \betavi \, \betavi^{\T} = B_i ,
\end{split}
\end{equation}
where $\langle X, Y \rangle  = \Tr (X^{\T} Y)$.
Since for any $\beta_{i} \in \Delta$
\begin{equation}
 \betavi \, \betavi^{\T} = B_i \ \Longleftrightarrow \ B_i e = \beta_i, \, B_i \succeq 0, \, {\rm rank}(B_i)=1, \, e^{\T} B_i e =1,
\end{equation}
a convex relaxation of \eqref{opt:mean-field-energy-min:separated:consensus:re} can be 
\begin{equation}	\label{opt:mean-field-energy-min:separated:consensus:relax}
\begin{split}
\min_{\betavi,\, \betav, \, B_i, \,B} \ \ &  -  \betavi^{\T} \ln \psivi + \langle \underline{\underline{M_{i}}} , B_{i} \rangle + \betavi^{\T} \ln \betavi \\
\mbox{s.t.} \ \  & \betavi = \betav \in \Delta, \\
			&  B_i e = \beta_i, \, e^{\T} B_i e =1, \, B_i = B \succeq 0, 
\end{split}
\end{equation}
where the rank constraint is not imposed and $B$ is a pseudo variable that all Agents share, i.e., it is a global variable that is required to be the same in the all decomposed optimizations.
The optimization \eqref{opt:mean-field-energy-min:separated:consensus:relax} provides a suboptimal solution for \eqref{opt:mean-field-energy-min:separated:consensus:re}
and the corresponding suboptimal value is a lower bound on the optimal value of \eqref{opt:mean-field-energy-min:separated:consensus:re}.
The resultant optimization \eqref{opt:mean-field-energy-min:separated:consensus:relax} is convex and
can be efficiently solved to find suboptimal solutions $\betavi^{\star} = \betav$ for all Agents $i \in V$.
In particular, we suggest to use dual decomposition methods (see Appendix \ref{subsec:dual-decomp}).

\subsubsection{Minimization of the Bethe free energy}

Consider the constrained minimization \eqref{opt:bethe-energy-min}.
This large-scale optimization over a graphical model can be decomposed into separated constrained minimizations for which Agent $i$ solves the optimization
\begin{equation}	\label{opt:bethe-energy-min:separated}
\begin{split}
\displaystyle
\min_{\betavi,\, \betamij} \ \
& - \betavi^{\T} \ln \psivi  - \sum_{j \in \Ncal(i)} [\betamij \circ \ln \psimij] \\
& \ 
(1- q_i) \betavk^{\T} \ln \betavi + \sum_{j \in \Ncal(i)}  [\betamij \circ \ln \betamij]
\\
\mbox{s.t.} \ \  & \betavi \in \Delta, \\
				& \betamij e = \betavi,\, j \in \Ncal(i)
\end{split}
\end{equation}
where the second constraint corresponds to the marginal probability constraint for the agents on edges connecting the node of Agent $i$.

Similar to the mean field energy approach, for fault detection and diagnosis with multiple hypotheses, assume that each agent has the same bank of hypothesized models and the objective of resultant distributed decision-making is to obtain optimal marginal and pairwise marginal beliefs $( \{\betavk\}_{k \in V}, \{\betamij\}_{(i,j) \in E} )$ that achieve the consistency in localized estimations, i.e.,
\begin{equation}	\label{eq:BeE:belief-consensus}
\begin{split}
\mbox{\bf Marginal Belief Consensus II: }
&
\beta_{i}(x) = \beta (x), \ \forall x \in \Xcal_{i}, \, \forall i \in V \\
&
\beta_{ij}(x,y) = b (x,y), \ \forall x \in \Xcal_{i},\, \forall y \in \Xcal_{j}, \, \forall (i,j) \in E 
\end{split}
\end{equation}
which can be rewritten as 
\begin{equation}	\label{eq:BeE:belief-consensus:vec}
\begin{split}
& \betavi = \betav , \, \forall i \in V, \, \mbox{ for some } \betav \in \Delta \\
& \betamij = B, \, \forall (i,j) \in E  \, \mbox{ for some } B \in  \Omega 
\end{split}
\end{equation}
where $\Omega \triangleq \{A \in \Real^{n \times n}_{+} : Ae = p \mbox{ and } e^{\T} A = q \mbox{ for some } p,q \in \Delta \}$.

The use of Bayesian hypothesis tests for FDD needs a special attention, for which the hypotheses at the nodes are homogeneous.
The pairwise marginal distributions $\{\betamij\}_{(i,j) \in E}$ are required to satisfy the conditions $\beta_{ij}(x,y) = 0$ for all $x \neq y$ for all $(i,j) \in E$,
which implies that the off-diagonal entries of $\betamij$ are zeros for all $(i,j) \in E$, or equivalently, the matrix $B$ in \eqref{eq:BeE:belief-consensus:vec} is a diagonal matrix. 

Incorporating the consensus requirement \eqref{eq:BeE:belief-consensus:vec} into \eqref{opt:bethe-energy-min:separated} results in a decomposed optimization
for which Agent $i$ solves
\begin{equation}	\label{opt:mean-field-energy-min:separated:consensus}
\begin{split}
\min_{\betavi,\, \betav} \ \ &  -  \betavi^{\T} \ln \psivi - \betavi^{\T} \underline{a_{i}} + \betavi^{\T} \ln \betavi \\
\mbox{s.t.} \ \  & \betavi = \betav \in \Delta,
\end{split}
\end{equation}
where $\underline{a_{i}} \triangleq \sum_{j \in \Ncal(i)} \ln ( {\rm diag} [\psimij] )$ and ${\rm diag}[A]$ denotes the vector whose elements are the diagonal entries of $A$ in order.
The resultant optimizations \eqref{opt:mean-field-energy-min:separated:consensus} are indeed convex
and can be efficiently solved to find global consensus optima $\betavi^{\star} = \betav$ for all Agents $i \in V$.
In particular, we suggest to use dual decomposition methods (see Appendix \ref{subsec:dual-decomp}).

\begin{remark}
In the aforementioned constrained optimization problems, the associated Lagrangian multipliers can be considered as prices of disagreement between agents (i.e., local beliefs).
The gradient dynamics of primal (the belief states) and dual variables (the prices of disagreement) should be explicitly written and interpreted in terms of
convergence rate, optimality, monotonicity, etc. 
\end{remark}
\section{Discussion}
\label{sec:discussion}

This section discusses several issues on the use of belief propagation for distributed statistical inference.
We also present some open questions that are not fully answered in this chapter.
The purpose of these discussions is to suggest future research directions for extensions and applications of BP/BO methods. 

\subsection{Open Problems}

For proper usage of belief propagation and optimization to tackle distributed statistical inference problems,
some underlying assumptions of BP/BO methods need to be further investigated.

\subsubsection{Correlated measurements}

\begin{figure}[t!]
\psfrag{A}[][]{Agent $1$}
\psfrag{B}[][]{Agent $2$}
\psfrag{C}[][]{Agent $3$}
\psfrag{E}[][]{E}
\psfrag{S}[][]{S}
\psfrag{P}[][]{P}
\psfrag{T}[][]{T}
\psfrag{R}[][]{R}
   \centering
   \includegraphics[width=0.68\textwidth]{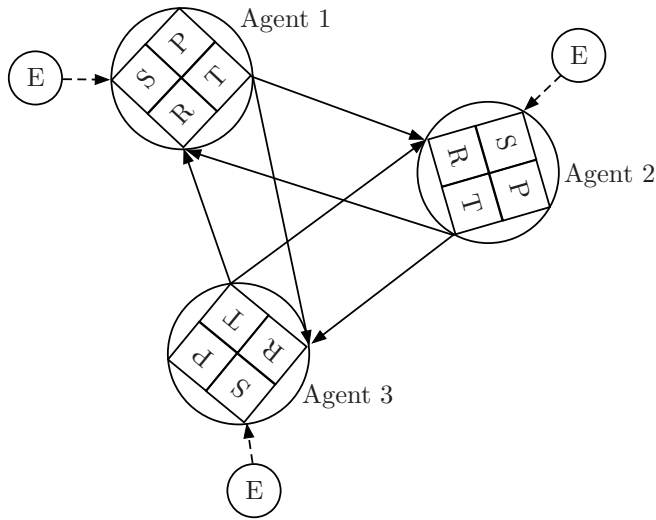} 
   \caption{A schematic cartoon of a Markov network for sensor fusion.
   The solid arrows correspond to communication links and the dotted arrows correspond to measurement mechanism.
   S: Sensor, P: Processor, S: Receiver, T: Transmitter, and E: Evidence (or Observational Event).}
   \label{fig:ex-sensor-fusion}
\end{figure}

Most of research works in the literature of belief propagation assume that each local measurement is conditionally independent given the other states at $V$ (even given the states at its neighborhood). 
In other words, the likelihood functions have the relations
\begin{equation}	\label{eq:cond-indep-bp}
p(y_k | x_k, x_{-k}) = p(y_k | x_k) \quad \forall k \in V .
\end{equation}
This assumption would be valid only for some special cases such as when the sensors are static (memoryless) and each source of uncertainty is localized.
To see the role of this assumption of conditional independence in belief propagation, consider the next example of sensor fusion.
\begin{example}	\label{ex:three-sensor-fusion}
Consider the Markov network model of sensor fusion depicted in Figure \ref{fig:ex-sensor-fusion}.
Messages from Agents $2$ and $3$ to Agent $1$ are computed by
\begin{equation} \label{ex:three-sensor-fusion:messages}
\mu_{j \to 1} (\x_1) \propto \sum_{\x_j \in \Xcal_j} p(\x_1 | \x_j) p (\x_j | y_j),\ \mbox{for } j =2,3 ,
\end{equation}
where $\{y_j\}$ are the local measurements that are available to Agents $j$.
Note that this is indeed a marginalization and results in
\begin{equation}
\mu_{j \to 1} (\x_1) \propto  p (\x_1 | y_j),\ \mbox{for } j =2,3 ,
\end{equation}
and the resultant belief is
\begin{equation}
\begin{split}
\beta_1(\x_1) 
& \propto p(\x_1|y_1) \mu_{2 \to 1} (\x_1) \mu_{3 \to 1} (\x_1) \\
& \propto p(\x_1|y_1) p(\x_1|y_2) p(\x_1|y_3)  .
\end{split}
\end{equation}
Under the assumption of conditional independence \eqref{eq:cond-indep-bp}, the belief can be rewritten as
\begin{equation} 	\label{ex:three-sensor-fusion:beliefs}
\begin{split}
\beta_1(\x_1) 
& \propto p(\x_1|y_1, y_2, y_3)
\end{split}
\end{equation}
that is the marginal probability of the state of Agent $1$ for given total measurements.
The marginal probabilities of Agents $2$ and $3$ can be computed in similar ways, viz.,
$\beta_2(\x_2)  \propto p(\x_2|y_1, y_2, y_3)$ and $\beta_3(\x_3)  \propto p(\x_3|y_1, y_2, y_3)$.
\end{example}
In the provious example,
notice that without assuming or guaranteeing conditional independence, the messages $\mu_{j \to i} (\x_i)$ for $i \neq j =1,2,3$ result in the beliefs $\beta_i(\x_i) \propto  \prod_{j=1}^{3} p(\x_i |y_j)$
which are not the same as the desired relations $\beta_i(\x_i) \propto p(\x_i |y_1, y_2, y_3)$.

Fortunately, for the case of homogeneous hypotheses in graphical models,
the likelihood functions \eqref{eq:cond-indep-bp} have the relations
\begin{equation}	\label{eq:cond-indep-bp:homo-hyp}
p(y_k | x_k, x_{-k}) = p(y_k | x_k) \prod_{j=1}^{n} \delta (x_k, x_j),
\quad \forall k \in V,
\end{equation}
where $\delta(x,y)$ refers to the standard scalar Dirac delta function.
This implies that for Example \ref{ex:three-sensor-fusion}, the message-passing algorithms \eqref{ex:three-sensor-fusion:messages} achieve the correct beliefs \eqref{ex:three-sensor-fusion:beliefs}
only if they satisfy the additional conditions of marginal belief consensus, viz., $\beta_{1}(\x) = \beta_{2}(\x)= \beta_{3}(\x)$ for all $\x \in \Xcal$.

%

\subsubsection{Pre vs. Post data processing and information fusion}

The primary goal of message-passing algorithms is to provide sufficient statistics for computations of marginal probabilities.
In the context of belief propagation, sufficient statistics of messages are properties that ensure the relations
\begin{equation}
\begin{split}
\beta_i \! \left( \x_{i} | y_i, \{\mu_{j \to i} \}_{j \in \Ncal(i)} \right) 
& = p \! \left( \x_{i} | Y = \{y_j\}_{j=1}^{n} \right) , \ \forall \x_{i} \in \Xcal_{i}, \ \forall i=1, \cdots, n.
\end{split}
\end{equation}
Message-passing algorithms can be considered as post data processing for information fusion,
whereas transmitting raw data, not subject to any data processing, is a naive method for computations of marginal probabilities.
Due to communication bandwidth limitations and cost of data storage, data transmission is not practical nor efficient.

In belief propagation algorithms based on graphical networked models, reducing communication costs has primary importance. 
Reducing size of transmitting messages with guaranteed exactness of resultant statistical inference is indeed to compute the smallest sufficient statistics.

\subsubsection{Suboptimality of consensus algorithms}

There was much research effort that studies convergence of message-passing algorithms in terms of properties of the graph $G=(V,E)$ (see \cite{alanyali2004distributed,olfati2006belief,moallemi2006consensus}, for example).
However, we should notice that convergence does not imply optimality in general.
Furthermore, such suboptimality can result in an arbitrarily bad decision whenever the estimation problem is connected to an optimal control problem,
in which inaccurate belief can deviate the resultant decision from an optimal one such that the achieved performance can be significantly worse off. 
For example, in~\cite{olfati2006belief}, the average-consensus algorithm and a belief propagation method are combined--such an algorithm was refereed to as belief consensus.
This belief consensus has many benefits such as scalability, convergence under varying network topology, etc.
However, it was regret that the authors did not provide any analysis of optimality and sub-optimality of their methods for distributed hypothesis tests.\footnote{Performance of a consensus algorithm can be arbitrarily bad--convergence vs. optimality.}
Notice that convergence or consensus of beliefs or messages does not necessarily imply optimality of the resultant hypothesis testing.


\subsection{MAP Consensus}

In Section~\ref{sec:belief-consensus:fdd-applications},
belief consensus constraints--conditions of \eqref{eq:MFE:belief-consensus} for the mean field energy minimization and conditions of \eqref{eq:BeE:belief-consensus} for the Bethe free energy minimization--are incorporated into belief optimization to reach agreement in marginal and pairwise marginal probability distributions of multiple hypotheses for given total measurements.

A popular statistical inference problem is to find a state that is the most probable from a probability distribution for given measurements.
For graphical models of distributed hypothesis testing, such a state can be obtained from m-MAP or j-MAP estimation.
Recall that an m-MAP estimator is a process to find state variables associated the nodes in a graphical model such that the corresponding marginal posterior probabilities have maximum values for given total measurements.
Similarly, but slightly differently, a j-MAP estimator is a process to find a configuration of state variables in a graphical model such that the corresponding joint posterior probability is a maximum for given total measurements.
For this purpose of inference, the aforementioned the max-product BP algorithms can be beneficial--using the max-product BP can reduce the communication costs, while the computational burdens of local processors would increase.

\section{Summary and Future Work}
\label{sec:conclusion}

This paper has developed methods of distributed Bayesian hypothesis testing, particularly, for applications to distributed fault detection and diagnosis in large-scale networked systems.
The presented methods are in the basis of belief propagation and optimization and use graphical models to represent the systems of consideration.
The resultant estimation problems reduce to solve distributed optimization for which the idea of belief optimization is adopted to use the concept of minimization of free energy to find an optimal probabilistic configuration of the state variables in Markov random fields.
For distributed computations of the associated constrained minimization problems, dual decomposition methods are used, which provide benefits of scalability and convergence.

Several discussions on issues of efficient and proper use of belief propagation and optimization for distributed statistical inference problems are provided.
Future research directions would be
(a) to develop further generalization of belief optimization using the concepts of region-based free energy representations--they are extensions of pairwise potential energy descriptions--and
(b) to evaluate exactness and compute approximation errors of an estimator that is obtained from minimizing an approximate free energy, to name of few.



\appendix

\section{Decomposition Algorithms}

\subsection{Iterative Dual Decomposition}
\label{subsec:dual-decomp}

This chapter primarily considers two problems. 
The first problem is a standard form of decomposable optimization with linear consistency (or complicating) constraints and
the second problem is its variation in which the local payoff functions are unequally weighted.

\begin{problem} 	\label{prob:separable_opt}
Consider an optimization with the separable payoff function, separable constraints, and equality consistency constraints of the form 
\begin{equation}	\label{eq:separable-cost-opt}
\begin{split}
\mbox{\rm maximize}\  & J(\x,\y) =  \sum_{k=1}^{N} \ell_{k} ( x_{k} , y_{k}) \\
\mbox{\rm subject to}\ & (x_{k}, y_{k}) \in \Fcal_{k}, \ k=1, \ldots, N, \\
				   & y_{k} =  C_{k} \z,\ k=1, \ldots, N, 	
\end{split}
\end{equation}
where $(x_{k}, y_{k})$ is the $k$th pair of separable decision variables that correspond to the separated {convex} cost functions $\ell_{k} ( x_{k} , y_{k})$,
$\Fcal_{k}$ denotes the $k$th constraint for the separated decision variable pair $(x_{k}, y_{k})$, and
$y_{k} =  C_{k} \z$ for $k=1, \ldots, N$ are the consistency constraints, which are the only coupled constraints over the separated decision variables. 
\end{problem}

\subsubsection{Lagrangian method and decomposition}

Consider the optimization~\eqref{eq:separable-cost-opt}.
An associated augmented Lagrangian to relax the consistency constraint is given by 
\begin{equation}	\label{eq:augmented-lagrangian}
\begin{split}
L(\x,\y,\z, \vb) &  =  \sum_{k=1}^{N} \ell_{k} ( x_{k} , y_{k}) - \sum_{k=1}^{N} \left\langle v_{k}, y_{k} - C_{k} \z \right\rangle\, ,  \\
		      & = \sum_{k=1}^{N} \left(  \ell_{k} ( x_{k} , y_{k}) -  \langle v_{k}, y_{k} \rangle \right) + \sum_{k=1}^{N}  \left\langle  v_{k}, C_{k}\z  \right\rangle\, , \\
		      & = \sum_{k=1}^{N} \left(  \ell_{k} ( x_{k} , y_{k}) -  \langle v_{k}, y_{k} \rangle \right) + \sum_{k=1}^{N}  \left\langle  C_{k}^{*} v_{k}, \z  \right\rangle\,  , \\
 \end{split}
\end{equation}
where $\x = [x_{1}, \cdots, x_{N}]^{\T}$, $\y = [y_{1}, \cdots, y_{N}]^{\T}$, $\vb = [v_{1}, \cdots, v_{N}]^{\T}$,\footnote{The bold refers to global variables while the non-bold refers to local variables.} and
the superscript refers to the associated adjoint operator.

Finding a saddle point that is a global optimal solution requires solving the two-stage optimization
\begin{equation}	\label{eq:two-stage_opt}
\inf_{\vb} \sup_{(\x,\y) \in \Fcal,\,\z} L(\x,\y,\z, \vb)\, , 
\end{equation}
where $\Fcal = \Fcal_{1} \times \cdots \times \Fcal_{N}$ refers to the product set of local (i.e., subsystem) constraints. 
From~\eqref{eq:augmented-lagrangian}, the optimization~\eqref{eq:two-stage_opt} can be rewritten as
\begin{equation}	\label{eq:two-stage_opt:rewritten}
\begin{array}{rl}
& \displaystyle  
\inf_{\vb}  \!
\underbrace{
\sup_{(\x,\y) \in \Fcal,\,\z} \! \left( \sum_{k=1}^{N} \left(  \ell_{k} ( x_{k} , y_{k}) -  \langle v_{k}, y_{k} \rangle \right) + \left\langle  C_{k}^{*} v_{k}, \z  \right\rangle \right) 
}_{
\left\{
\begin{array}{l}
\displaystyle 
\inf_{\vb} \sup_{(\x,\y) \in \Fcal} \left( \sum_{k=1}^{N} \left(  \ell_{k} ( x_{k} , y_{k}) -  \langle v_{k}, y_{k} \rangle \right) \right) \\ \ \ \ \ \ \ \ \ \ \  \mbox{if } \displaystyle  \C^* \vb = 0 \\
+ \infty  \ \ \ \ \    \mbox{otherwise}
\end{array}
\right.}\, , 
\\
& =
\displaystyle 
\inf_{\C^{\T} \vb = 0} \sup_{(\x,\y) \in \Fcal} \left( \sum_{k=1}^{N} \left(  \ell_{k} ( x_{k} , y_{k}) -  \langle v_{k}, y_{k} \rangle \right) \right) \, , 
\\
& =
\displaystyle 
\inf_{\C^{\T} \vb = 0}  \sum_{k=1}^{N} \left(  \sup_{(x_{k},y_{k}) \in \Fcal_{k}} \left(  \ell_{k} ( x_{k} , y_{k}) -  \langle v_{k}, y_{k} \rangle \right) \right) \, , 
\end{array}
\end{equation}
where $\C^{\T} = [C_{1}^{\T}, \cdots, C_{N}^{\T}]$.

The optimization~\eqref{eq:two-stage_opt} can be decomposed into two convex programs:
\begin{equation}	\label{eq:slave-problem}
\mbox{\bf Slave Problem: }
\sup_{(x_{k},y_{k}) \in \Fcal_k} \underbrace{\left(  \ell_{k} ( x_{k} , y_{k}) -  \langle v_{k}, y_{k} \rangle \right)}_{S_{k} (x_{k}, y_{k} | v_{k})}\, ,
\end{equation}
for $k = 1, \ldots, N$,
and
\begin{equation}	\label{eq:master-problem}
\mbox{\bf Master Problem: }
\inf_{\C^{\T} \vb = 0}  \sum_{k=1}^{N} \underbrace{S_{k} (x^{\opt}_{k}, y^{\opt}_{k} | v_{k})}_{Q_{k}(v_{k})}\, , 
\end{equation}
where $(x^{\opt}_{k}, y^{\opt}_{k})$ refers to the optimal solution pair for the Slave Problem~\eqref{eq:slave-problem} for given $v_{k}$.


%
\begin{figure}[b!]
\psfrag{M}[][]{$\begin{array}{c} \mbox{\sc Master} \\ \mbox{\sc Processor} \end{array}$}
\psfrag{A}[][]{$\small  \begin{array}{c}\mbox{\rm Local} \\ \mbox{\rm Agent} \\1\end{array}$}
\psfrag{B}[][]{$\small \begin{array}{c}\mbox{\rm Local} \\ \mbox{\rm Agent} \\ 2\end{array}$}
\psfrag{C}[][]{$\small \begin{array}{c}\mbox{\rm Local} \\ \mbox{\rm Agent} \\ N-1\end{array}$}
\psfrag{D}[][]{$\small \begin{array}{c}\mbox{\rm Local} \\ \mbox{\rm Agent} \\ N\end{array}$}
\psfrag{S}[][]{$\cdots$}
\psfrag{a}[][][0.75]{$g_{1}^{(n)}$}
\psfrag{b}[][][0.75]{$g_{2}^{(n)}$}
\psfrag{c}[][][0.75]{$g_{N-1}^{(n)}$}
\psfrag{d}[][][0.75]{$g_{N}^{(n)}$}
\psfrag{e}[][][0.75]{$v_{1}^{(n)}$}
\psfrag{f}[][][0.75]{$v_{2}^{(n)}$}
\psfrag{g}[][][0.75]{$v_{N-1}^{(n)}$}
\psfrag{h}[][][0.75]{$v_{N}^{(n)}$}
   \centering
   \includegraphics[width=0.68\textwidth]{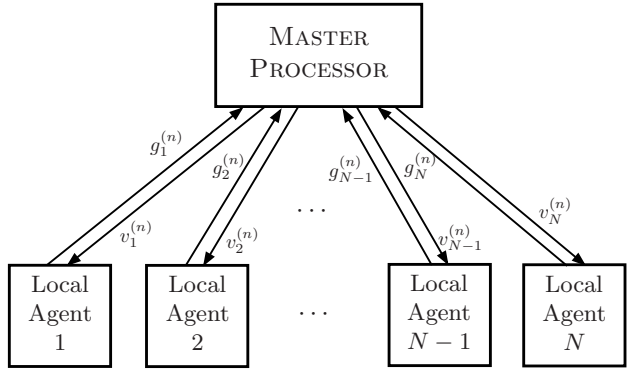} 
   \caption{Iterative dual decomposition of sequentially reporting public variables $\{g_{k}^{(n)}\}$ and assigning prices $\{v_{k}^{(n)}\}$. The superscript $(n)$ refers to the iteration sequence.}
   \label{fig:dual-decomposition}
\end{figure} 
%


\subsubsection{Projection-(Sub)gradient method}

The Master Problem~\eqref{eq:master-problem} can be solved using a first-order (sub-)gradient projection method,
whereas the Slave Problem~\eqref{eq:slave-problem} is of much smaller size and can be accurately and efficiently solved by a second-order method such as an interior-point algorithm~\cite{Boyd_CVX:book,Nesterov:book}.
Consider that the Master Problem~\eqref{eq:master-problem} can be rewritten as
\begin{equation}	\label{eq:master-problem:2}
\inf_{\C^{\T} \vb = 0}  \sum_{k=1}^{N} Q_{k}(v_{k})\, , 
\end{equation}
where $Q_{k}(v_{k})$ is convex in $v_{k}$ for all $k=1, \ldots, N$.
Define a linear subspace $\Mcal \triangleq \{\vb \in \Real^{\bullet} :  \C^{\T} \vb = 0 \}$, which is the null-space of the matrix $\C^{\T}$.
The optimization (\ref{eq:master-problem:2}) can be solved using a subgradient-projection method:
\begin{equation}	\label{eq:subgradient-projection}
\vb^{(n+1)} : = \Pcal_{\Mcal} (\vb^{(n)} - \alpha_{n} g^{(n)} (\vb^{(n)}))\, ,
\end{equation}
where $\Pcal_{\Mcal} : \Real^{\bullet} \rightarrow \Mcal$ refers to the projection on the subspace $\Mcal$, $g^{(n)} : \Real^{\bullet}  \rightarrow \Real^{\bullet}$ denoted the subgradient, i.e., $g \in \partial_{\vb} \sum Q_{k} (v_{k})$,
and $\alpha_{n}$ is a step size that can be selected in any of standard ways (e.g., constant, diminishing, etc.).
The sub-differential can be represented as 
\begin{equation}
\begin{split}
 \partial_{\vb} \left( \sum_{k=1}^{N} Q_{k} (v_{k}) \right) &=  \partial_{v_{1}} Q_{1} (v_{1}) \times \cdots \times \partial_{v_{N}} Q_{N} (v_{N})\ ,  \\
 										    & = \{ -\y^{\opt} \} \, , 
\end{split}
\end{equation}
where $\y^{\opt}$ denotes the concatenation of optimal solutions of the Slave Problem~\eqref{eq:slave-problem} for a given sequence $\{v_{k}\}$.
In other words, local subsystems are required to sequentially report the computed public variables to the supervisor (or price-planner).
Therefore, the update rule for the subgradient-projection method~\eqref{eq:subgradient-projection} can be rewritten as
\begin{equation}	\label{eq:subgradient-projection:2}
\vb^{(n+1)} : = \Pcal_{\Mcal}\!\! \left( \vb^{(n)} +  \alpha_{n} \y^{(n)}  \right) \, , 
\end{equation}
where the superscript $\opt$ of $\y$ is removed for notational convenience.
Furthermore, it is not hard to see that 
\begin{equation}
\Pcal_{\Mcal}(z) = \left( \Id - C(C^{\T} C)^{-1}C^{\T} \right) z \, ,
\end{equation}
so that 
\begin{equation}	\label{eq:subgradient-projection:3}
\begin{split}
\vb^{(n+1)} & := \left( \Id - C(C^{\T} C)^{-1}C^{\T} \right)  \left( \vb^{(n)} +  \alpha_{n} \y^{(n)}  \right) \,  , \\
		  & :=  \vb^{(n)} + \alpha_{n} \underbrace{\left( \Id - C(C^{\T} C)^{-1}C^{\T} \right)}_{U} \y^{(n)} \, , 
\end{split}
\end{equation}
where the computation of the matrix $U$ needs to be performed only once and can be done offline (before performing optimization).


\subsubsection{Separable cost with coupled inequalities}

\begin{problem}
Consider an optimization with the separable payoff function, separable constraints, and coupled inequality constraints of the form
\begin{equation}	\label{eq:separable-cost-opt:coupled-ineq-constraints}
\begin{split}
\mbox{\rm maximize}\  & J(\x) =  \sum_{k=1}^{N} \ell_{k} ( x_{k}) \\
\mbox{\rm subject to}\ & x_{k}  \in \Fcal_{k}, \ k=1, \ldots, N  \, , \\
				   &  \C \x \geq 0	\, , 
\end{split}
\end{equation}
where $x_{k}$ is the $k$th separable decision variable that corresponds to the separated {\it convex} cost functions $\ell_{k} ( x_{k})$,
$\Fcal_{k}$ denotes the $k$th constraint for $x_{k}$, and
$\C \x =  \sum_{k=1}^{N} C_{k} x_{k}$ for $k=1, \ldots, N$ are coupled inequality constraints. 
\end{problem}

An associated augmented Lagrangian is 
\begin{equation}	\label{eq:augmented-lagrangian:coupled-ineq-constraint}
\begin{split}
L(\x, \vb) &  =  \sum_{k=1}^{N} \ell_{k} ( x_{k} ) -  \left\langle \vb ,  \C \x  \right\rangle \, , \\
	       &  =  \sum_{k=1}^{N} \ell_{k} ( x_{k} ) - \sum_{k=1}^{N} \left\langle \vb ,  C_{k} x_{k} \right\rangle \, ,  \\
 \end{split}
\end{equation}
where $\vb \geq 0$.
The constrained optimization \eqref{eq:separable-cost-opt:coupled-ineq-constraints} can be decomposed into the two-stage optimization
\begin{equation}	\label{eq:two-stage_opt:coupled-ineq-constraint}
\begin{split}
\inf_{\vb \geq 0} &  \sup_{\x \in \Fcal}   L(\x, \vb)  \\
=
&
\inf_{\vb \geq 0}  \left( \sup_{\x \in \Fcal} \left(  \sum_{k=1}^{N} \ell_{k} ( x_{k} ) - \sum_{k=1}^{N} \left\langle \vb ,  C_{k} x_{k} \right\rangle  \right) \right) \, , \\
=
&
\inf_{\vb \geq 0} \left(  \sum_{k=1}^{N} \underbrace{ \sup_{x_{k} \in \Fcal_{k}}   \left(  \ell_{k} ( x_{k} ) - \left\langle \vb ,  C_{k} x_{k} \right\rangle  \right)}_{Q_{k}(\vb)}  \right)\, ,  \\
\end{split}
\end{equation}
where $\Fcal = \Fcal_{1} \times \cdots \times \Fcal_{N}$ refers to the product set of local (i.e., subsystem) constraints. 
The optimization~\eqref{eq:two-stage_opt:coupled-ineq-constraint} can be decomposed into two convex programs:
\begin{equation}	\label{eq:slave-problem:coupled-ineq-constraint}
\mbox{\bf Slave Problem: }
\sup_{x_{k} \in \Fcal_{k}} \underbrace{\left(  \ell_{k} ( x_{k} ) -  \langle \vb, C_{k} x_{k} \rangle \right)}_{S_{k} (x_{k} | \vb)} \, , 
\end{equation}
for $k = 1, \ldots, N$,
and
\begin{equation}	\label{eq:master-problem:coupled-ineq-constraint}
\mbox{\bf Master Problem: }
\inf_{\vb \geq 0}  \sum_{k=1}^{N} \underbrace{S_{k} ( x^{\opt}_{k} | \vb )}_{Q_{k}(\vb)} \, , 
\end{equation}
where $x^{\opt}_{k}$ refers to the optimal solution pair for the Slave Problem~\eqref{eq:slave-problem:coupled-ineq-constraint} for given $\vb$.
A similar projection-subgradient method as aforementioned can be used to solve this problem.\\
{\bf Projection-(Sub)gradient Method: }
Starting from a feasible dual variable $\vb^{(0)} \geq 0$, the sequences of primal-dual solutions can be computed as follows:
\begin{equation}
x_{k}^{(n)} : =
\arg \max_{x_{k} \in \Fcal_{k}} \left(  \ell_{k} ( x_{k} ) -  \langle \vb^{(n)} , C_{k} x_{k} \rangle \right) \, ,
\end{equation}
and
\begin{equation}
\vb^{(n+1)} : = 
\left( \vb^{(n)} + \alpha_{n} \sum_{k=1}^{N} C_{k} x_{k}^{(n)}  \right)_{\!\!+} \, , 
\end{equation}
where $(a)_{+}$ has the $i$th element defined as $a_{i}$ if $a_{i} \geq 0$ and $0$ otherwise.

\bibliographystyle{abbrv}
\bibliography{belief-propagation}

\end{document}